\documentclass[a4paper,11pt]{llncs}
\usepackage{fullpage,graphicx}

\newcommand{\Oh}[1]
    {\ensuremath{\mathcal{O}\!\left( {#1} \right)}}
\newcommand{\occ}
    {\ensuremath{\mathit{occ}}}
\newcommand{\heap}
    {\ensuremath{\mathit{Heap}}}
\newcommand{\ST}
    {\ensuremath{\mathit{ST}}}
\newcommand{\SA}
    {\ensuremath{\mathit{SA}}}
\newcommand{\sheap}
    {\ensuremath{\mathit{S\mbox{-}Heap}}}
\newcommand{\build}
    {\ensuremath{\mathit{Build}}}

\begin{document}

\title{New Algorithms for Position Heaps}
\author{Travis Gagie\inst{1} \and Wing-Kai Hon\inst{2} \and Tsung-Han Ku\inst{2}}
\institute{Department of Computer Science, University of Helsinki, Finland\\
    \and Department of Computer Science, National Tsing Hua University, Taiwan}
\maketitle

\begin{abstract}
We present several results about position heaps, a relatively new alternative to suffix trees and suffix arrays.  First, we show that, if we limit the maximum length of patterns to be sought, then we can also limit the height of the heap and reduce the worst-case cost of insertions and deletions.  Second, we show how to build a position heap in linear time independent of the size of the alphabet.  Third, we show how to augment a position heap such that it supports access to the corresponding suffix array, and vice versa.  Fourth, we introduce a variant of a position heap that can be simulated efficiently by a compressed suffix array with a linear number of extra bits.
\end{abstract}

\section{Introduction} \label{sec:intro}

String-indexing data structure have played a central role in pattern matching at least since the introduction of suffix trees forty years ago, and their importance has only increased with the introduction of suffix arrays, compressed suffix arrays, FM-indexes, etc.  There are are still many open problems about them, however, such as how best to make them dynamic.  There are now fairly practical dynamic versions of suffix arrays and FM-indexes but these have poor worst-case theoretical bounds for updates.  Relatively recently, Ehrenfeucht, McConnell, Osheim and Woo~\cite{EMOW11} introduced a new and simple indexing data structure, called a position-heap, and showed how it easily can be made dynamic (albeit with a logarithmic slowdown for searches and also with a poor worst-case bound for updates).  Like suffix trees and suffix arrays, position heaps take linear space and supports searching in time proportional to the length of the pattern plus the number number of occurrences reported, which is optimal.  Ehrehfeucht et al. gave a construction algorithm that works in linear time when the size of the alphabet is constant.  Shortly thereafter, Kucherov~\cite{Kuc??} gave a simpler, online construction that also takes linear time when the alphabet size is constant.  Ehrenfeucht et al.'s and Kucherov's constructions of position heaps are analogous to Weiner's~\cite{Wei73} and Ukkonen's~\cite{Ukk95} construction of suffix trees, respectively, and Kucherov asked whether there is a construction that takes linear time independent of the alphabet size, analogous to Farach's~\cite{Far97} construction of suffix trees.  Kucherov also asked whether position heaps can be compressed, as can suffix trees, suffix arrays and FM-indexes.  Most recently, Nakashima, I, Inenaga, Bannai and Takeda~\cite{NIIBT12} showed how to build the position heap for a set of strings given as a trie in linear time when the alphabet size is constant.

In this paper we answer some of the open problems about position heaps.  We show in Section~\ref{sec:limiting} that, if we limit the maximum length of patterns to be sought, then we can use a position heap with limited height as an index, which reduces the maximum cost of updating the heap after we make insertions or deletions in the string.  In many practical applications we are interested only in fairly short patterns anyway, so this seems like a reasonable tradeoff.  We also note in that, if we replace a splay tree by an AVL-tree in Ehrenfeucht et al.'s implementation of dynamic position heaps, then their time bounds become worst-case instead of amortized.  In Section~\ref{sec:tree2heap} we show how to turn a suffix tree into a position heap in linear time independent of the alphabet size, using a simple modification of a recent algorithm by Bannai, Inenaga and Takeda~\cite{BIT12} for building the LZ78 parse from a straight-line program.  Combined with Farach's algorithm for building suffix trees in linear time, this means we can build position heaps in linear time independent of the alphabet size, answering Kucherov's first question affirmatively.  In Section~\ref{sec:heap2array} we show how a to augment a position heap with $\Oh{n \log h}$ bits such that it supports $\Oh{1}$-time access to the corresponding suffix array and inverse suffix array, where $n$ is the length of the string and $h$ is the height of the heap.  Ehrenfeucht et al. showed that, although $h$ can be as large as $n$ in the worst case, it is typically $\Oh{\log n}$.  We also show how to augment a compressed suffix array with $\Oh{n \log h}$ bits such that it supports access to the position heap in the same time needed to access the suffix array and inverse suffix array.  Finally, in Section~\ref{sec:sheap} we introduce a variant of a position heap, which we call a suffix heap, that still supports indexed pattern matching but which can be simulated by a compressed suffix array with only a linear number of extra bits.  This seems at least partly to answer Kucherov's second question affirmatively as well.

\section{Position Heaps} \label{sec:heaps}

\begin{figure}[t]
\begin{center}
\includegraphics[width=40ex]{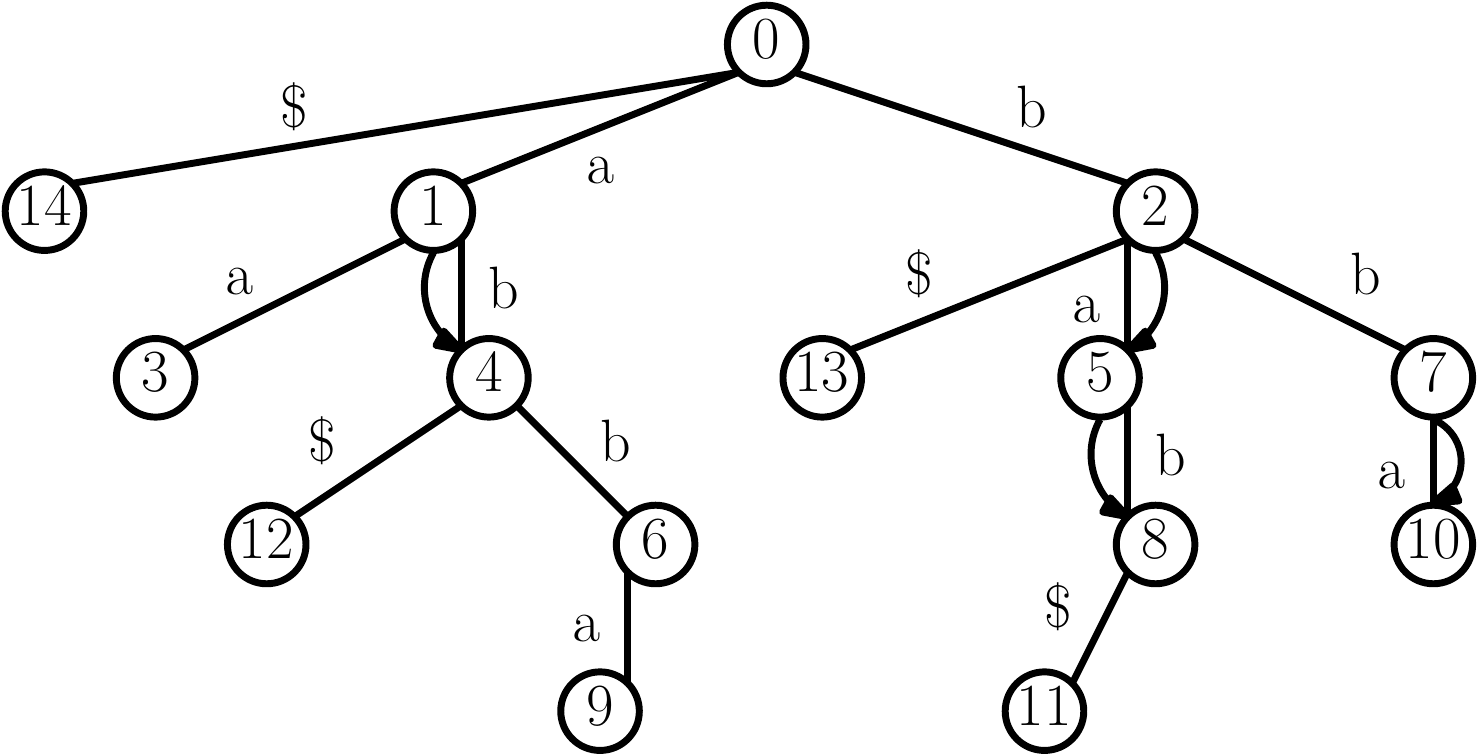}
\caption{The position heap $\heap$ for \(S = \mathrm{abaababbabbab\$}\).}
\label{fig:heap}
\end{center}
\end{figure}

Ehrenfeucht et al.'s position heap data structure is a modification of an older data structure by Coffman and Eve~\cite{CE70} for hashing.  Kucherov gave a simplified definition according to which, for a string \(S [1..n]\) terminated by a special symbol \(S [n] = \mathrm{\$}\), the position heap is the trie $\heap$ in which
\begin{itemize}
\item the root is labelled 0 and the other nodes are labelled 1 to n such that parents' labels are greater than their children's labels;
\item for \(1 \leq i \leq n\), the path label of the node labelled $i$ is a prefix of \(S [i..n]\);
\item for \(1 \leq i \leq n\), the node labelled $i$ stores a pointer (called its maximal-reach pointer) to the deepest node whose path label is a prefix of \(S [i..n]\).
\end{itemize}
For example, if \(S = \mathrm{abaababbabbab\$}\) then $\heap$ is as shown in Figure~\ref{fig:heap} (except that maximal-reach pointers are omitted there when they point back to the nodes themselves).  One reason to label the root 0 is so that, for \(1 \leq i \leq n\), \(S [i]\) is equal to the first edge label on the path from the root to the node labelled $i$.

To be able to use $\heap$ for indexed pattern matching in $S$ we store, first, an array of pointers such that, given $i$, in $\Oh{1}$ time we can find the node labelled $i$; and second, a data structure such that, given $i$ and $j$, in $\Oh{1}$ time we can determine whether the node labelled $i$ is an ancestor of the node labelled $j$.  The total space for $\heap$ and these data structures is $\Oh{n \log n}$ bits, i.e., linear space on a word RAM.

To search for a pattern \(P [1..m]\) in $S$, we start at the root and descend to the deepest node $v$ whose path label is a prefix of $P$.  If the $v$'s depth \(d = m\), then we report the label of each node that is either in the subtree of $v$ or on the path from the root to $v$ with a maximal reach pointer into the subtree of $v$.  Otherwise, we build a list containing the label of each node on the path from the root to $v$ with a maximal reach pointer to $v$.  We return to the root and descend to the deepest node $v'$ whose path label is a prefix of \(P [d + 1..m]\).  If the depth $d'$ of $v'$ is \(m - d\), then we report each label $i$ in our list for which the node labelled \(i + d\) is either in the subtree of $v'$ or on the path from the root to $v'$ with a maximal-reach pointer into the subtree of $v'$. Otherwise, we filter our list, keeping each label $i$ only if the node labelled \(i + d\) is on the path from the root to $v'$ with a maximal reach pointer to $v'$.  We return to the root and descend again, using $d'$ in place of $d$, and keep repeatedly descending until we reach the end of $P$.  By induction, this yields a list of the starting positions of the occurrences of $P$ in $S$ and, with the data structures mentioned above, takes time linear in $m$ and the number of those occurrences.

For example, to search for \(P = \mathrm{aabab}\) in \(S = \mathrm{abaababbabbab\$}\), we start at the root and descend along two edges labelled \(P [1] = P [2] = \mathrm{a}\) to the node $v$ labelled 3.  Since $v$ is at depth only \(d = 2 \leq m = 5\), we check the nodes labelled 1 and 3 and then, since the former's maximal-reach pointer is not to the latter, build a list containing only 3.  We return to the root and descend along edges labelled \(P [3] = \mathrm{b}\), \(P [4] = \mathrm{a}\) and \(P [5] = \mathrm{b}\) to the node $v'$ labelled 8.  Since $v'$ is at depth \(3 = m - d\), we find the node labelled \(3 + d = 5\) and, since it is on the path from the root to the $v'$ and its maximal-reach pointer is into the subtree of $v'$, we report position 3.

\section{Limiting Length and Height} \label{sec:limiting}

If we will never search for a pattern of length greater than $M$, then we can easily build a position heap of height $\Oh{M}$ that works as an index for $S$.  To do this, we make two copies of $S$ called $S'$ and $S''$; insert a unique character after every \(2 M\) characters, counting from the first character of $S'$ and the \((M + 1)\)st character of $S''$; and build the position heap for \(S'\,!\,S''\), where $!$ is another unique character.  We refer to the inserted unique characters and $!$ as dividers and to the substrings of $S'$ and $S''$ strictly between dividers as blocks.  For example, if \(S = \mathrm{abaababbabbab\$}\) and \(M = 3\), then \(S' = \mathrm{abaaba\,\#_1\,bbabba\,\#_2\,b\$}\), \(S'' = \mathrm{ababba\,\#_3\,bbab\$}\) and we build the position heap for
\[S'\,!\,S'' = \mathrm{abaaba\,\#_1\,bbabba\,\#_2\,b\$\,!\,ababba\,\#_3\,bbab\$}\,.\]

Notice that any substring of $S$ with length at most $M$ occurs in either $S'$ or $S''$ or both.  Moreover, given the endpoints of a substring in \(S'\,!\,S''\), in $\Oh{1}$ time we can determine whether it contains any dividers and, if not, where it occurs in $S$.  Therefore, we can use the position heap for \(S'\,!\,S''\) as an index for $S$.  The position heap for \(S'\,!\,S''\) has height at most a factor of 2 larger than the height of the position heap for $S$ and the dividers guarantee there are no common prefixes in $S$ longer than \(2 M\), so the position heap for \(S'\,!\,S''\) has height $\Oh{M}$.

If we insert or delete a substring in $S$, then we should update $S'$ and $S''$ to maintain the invariants that every substring of $S$ with length at most $M$ occurs in either $S'$ or $S''$ or both, and that the position heap for \(S'\,!\,S''\) has height $\Oh{M}$.  Consider first how we update $S'$ when we insert a substring of length at most \(4 M\) into $S$.  We insert that substring into the appropriate block of $S'$; if that block then has length more than \(4 M\), then we split the block into two parts, each of length between \(2 M\) and \(4 M\), and insert a new divider between them.  If we insert a a substring with length greater than \(4 M\) into $S$, then we split that substring into blocks of length at most \(2 M\) separated by dividers, split the block of $S'$ where the substring is to be inserted into two parts, concatenate the first part with the first block of the substring and concatenate the last block of the substring with the second part.

If we delete a substring of $S$, then we delete any blocks of $S'$ completely contained in that substring, then perform separate deletions from the blocks where the substring starts and finishes.  To delete a substring from a single block of $S'$, we delete that substring and then check whether the block still has length at least \(2 M\).  If not, we remove the divider between that block and an adjacent one (assuming $S$ is still long enough for there to be another block); if the resulting block then has length more than \(4 M\), then we split it into two parts, each with length between \(2 M\) and \(4 M\), and insert a new divider between them.

Once we have updated $S'$, we update $S''$ so that the blocks of $S''$ are again centered on the dividers in $S'$ and have length exactly \(2 M\) (or less if they reach an end of $S$).  Notice that inserting or deleting a substring of length $\ell$ into or from $S$ requires inserting or deleting $\Oh{1}$ substrings of length $\Oh{\ell}$ into or from \(S'\,!\,S''\).  For example, if \(S = \mathrm{abaababbabbab\$}\), \(M = 3\) and we insert \(\mathrm{bba}\) in position 5 to obtain \(S = \mathrm{abaa} \mathit{bba} \mathrm{babbabbab\$}\), then we update $S'$ to be \(\mathrm{abaa} \mathit{bba} \mathrm{ba\,\#_1\,bbabba\,\#_2\,b\$}\) and $S''$ to be \(\mathit{a} \mathrm{babba\,\#_3\,bbab\$}\).

Ehrenfeucht et al.'s dynamic index has two parts, a dynamic position heap and the data structure for storing the dynamic string itself.  They suggested using a splay tree to store the dynamic string but noted that this choice gives only amortized time bounds.  If we use their dynamic position heap for \(S'\,!\,S''\) exactly as they described but use an AVL-tree (which can also be split and joined in logarithmic time) instead of a splay tree to store \(S'\,!\,S''\), then we obtain the following result with no amortization.  We will give more details in the full version of this paper.  Our use of dividers makes the alphabet size more than constant but, as we show in the next Section, it is still possible to build the position heap in linear time.

\begin{theorem} \label{thm:limited}
If we will never search for a pattern of length greater than $M$ in a dynamic string $S$, then we can maintain a position heap that works as an index for $S$ such that
\begin{itemize}
\item searching for a pattern of length \(m \leq M\) takes $\Oh{m \log |S| + \occ}$ time,
\item inserting a substring of length $\ell$ takes $\Oh{(M + \ell) M \log (|S| + \ell)}$ time, 
\item deleting a substring of length $\ell$ takes $\Oh{(M + \ell) M \log |S|}$ time.
\end{itemize}
\end{theorem}

\section{Turning a Suffix Tree into a Position Heap} \label{sec:tree2heap}

Bannai et al. recently gave an algorithm for computing the LZ78 parse of a string from a straight-line program for that string.  A key idea in their algorithm is to build the LZ78 trie superimposed on the suffix tree for the string.  To compute the LZ78 parse normally, we start at the left with an empty dictionary; at each step, we take as the next phrase the shortest prefix of the remainder of the string, that is not yet in the dictionary; we add that phrase to the dictionary and delete it from the beginning of the remainder of the string.  The trie of the phrases in the dictionary when we finish parsing is the LZ78 trie.  If we delete only the first character of the remainder of the string at each step, instead, then the trie of the phrases when we finish parsing is the position heap.  In this section we use this idea to turn a suffix tree into a position heap in linear time independent of the alphabet size.

\begin{figure}[t]
\begin{center}
\includegraphics[width=50ex]{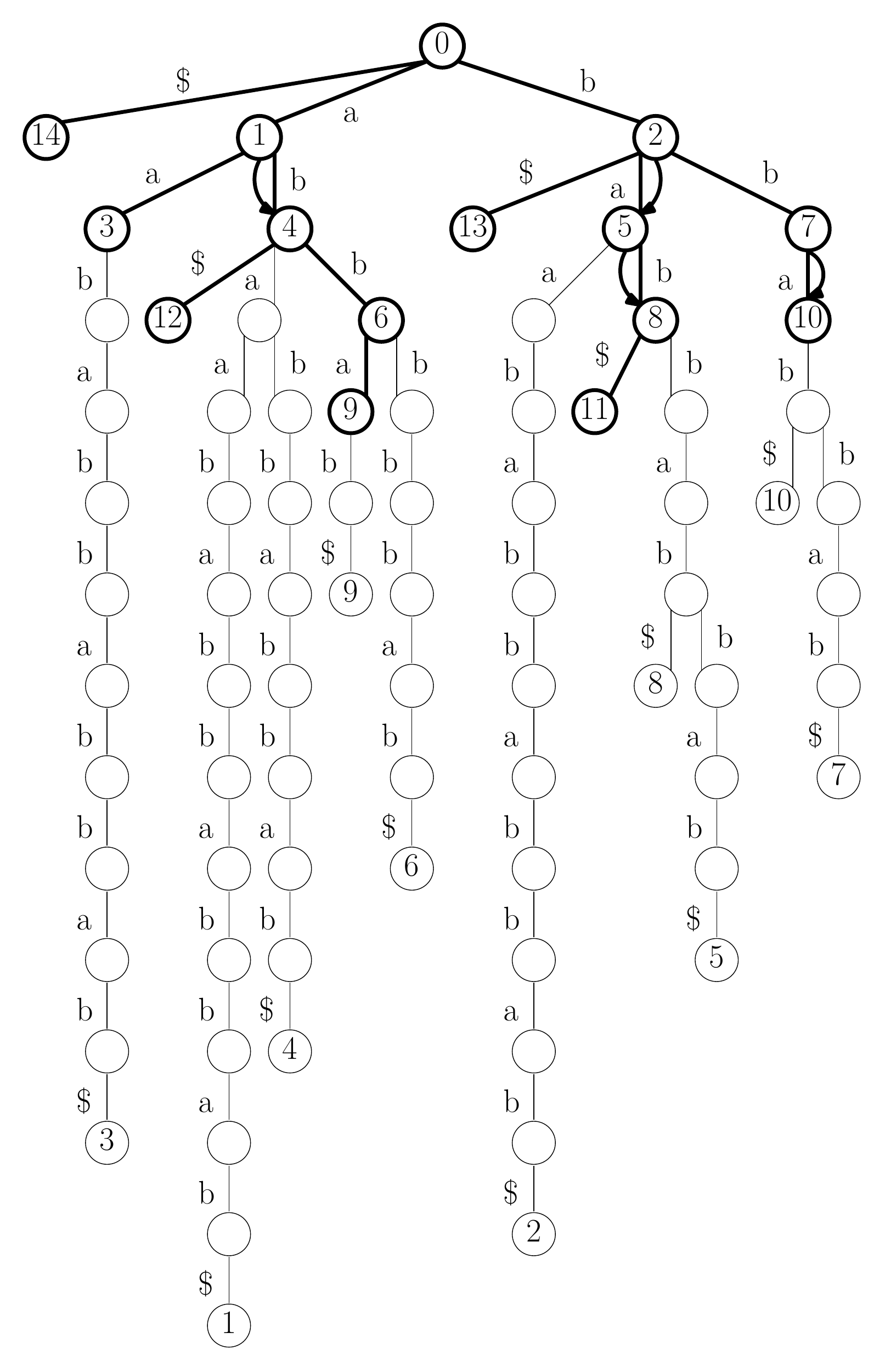}
\caption{The position heap $\heap$ (in heavy lines) overlaid on the suffix trie for \(S = \mathrm{abaababbabbab\$}\).}
\label{fig:trie}
\end{center}
\end{figure}

A simple way to build $\heap$ is to build the suffix trie for $S$ (i.e., the trie of all its suffixes); label each leaf with the starting position of the suffix which is its path label; label the root 0; for \(1 \leq i \leq n\), move each leaf's label to its highest unlabelled ancestor (or, if there are no unlabelled ancestors, leave the label on the leaf); and finally, for \(1 \leq i \leq n\), add a maximal-reach pointer from the node labelled $i$ to the deepest labelled ancestor of the leaf originally labelled $i$.  The correctness of this algorithm follows from the definition of the position heap.  Figure~\ref{fig:trie} shows $\heap$ overlaid on the suffix trie for \(S = \mathrm{abaababbabbab\$}\).

Suppose we already have built and preprocessed the suffix trie of $S$ such that in $\Oh{1}$ time, first, we can mark nodes; second, given a node, we can find its lowest marked ancestor; and third, given a node and a depth, we can find that node's ancestor at that depth.  Then we can perform the algorithm we have just described in linear time independent of the alphabet size, marking nodes whenever we move a label to them.  Since the suffix trie has size \(\Theta (n^2)\), however, building it explicitly and preprocessing it takes \(\Omega (n^2)\) time.

Bannai et al. showed how we can use the suffix tree $\ST$ for $S$ as a representation of the suffix trie.  Suppose we have already built two copies $\ST_1$ and $\ST_2$ of $\ST$ with the same nodes.  Westbrook~\cite{Wes92} (see also~\cite{AFILS95}) showed how we can preprocess $\ST_1$ in linear time such that in $\Oh{1}$ amortized time, first, we can mark nodes; second, given a node, we can find its lowest marked ancestor; and third, we can insert a node in the middle of an edge.  Berkman and Vishkin~\cite{BV94} (see also~\cite{BF04}) showed how we can preprocess $\ST_2$ in linear time such that, given a node and a depth, we can find that node's ancestor at that depth in $\Oh{1}$ time.  We work in $\ST_1$, which is dynamic; $\ST_2$ remains static.

\begin{figure}[t]
\begin{center}
\includegraphics[width=50ex]{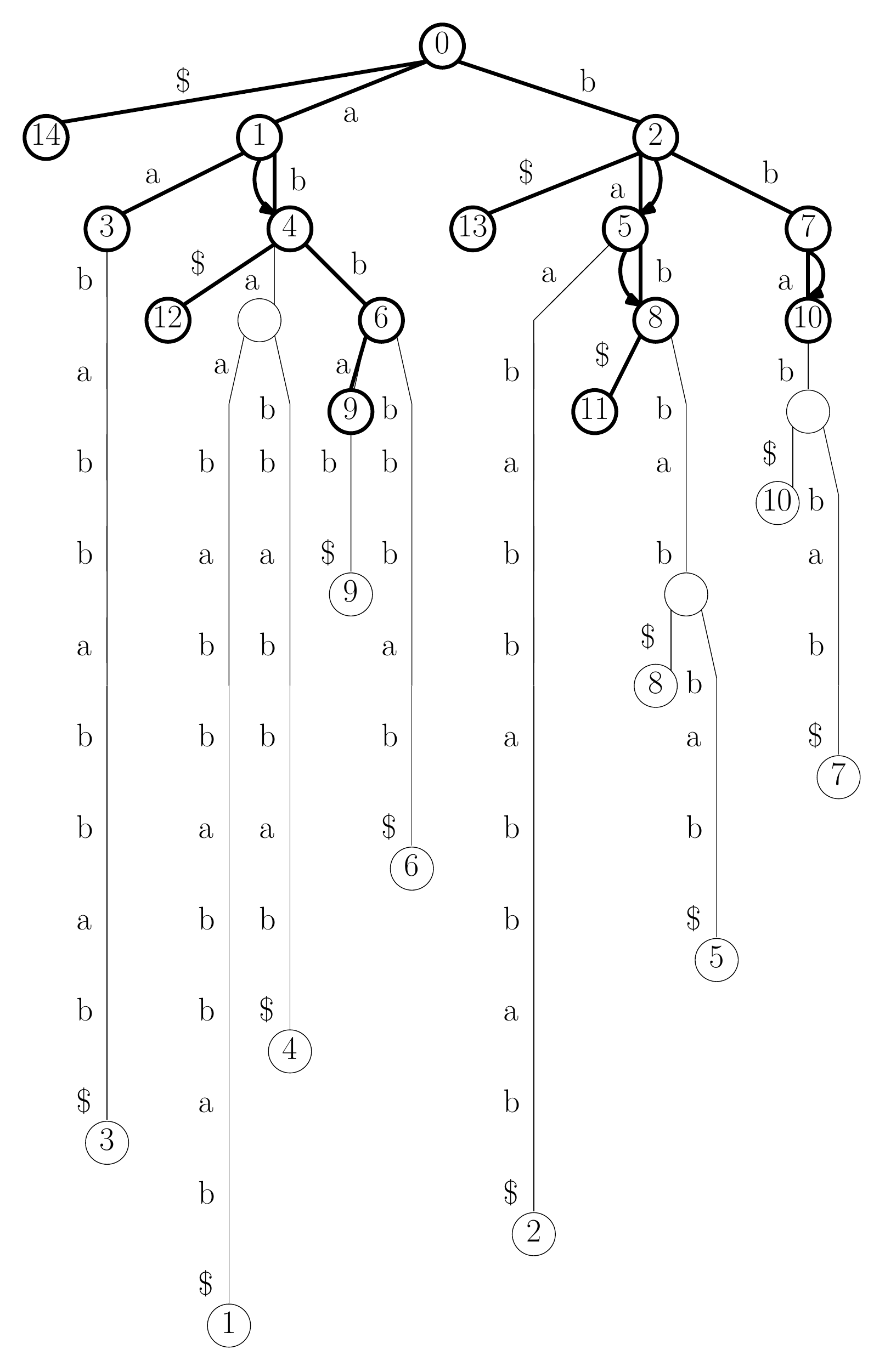}
\caption{The position heap $\heap$ (in heavy lines) overlaid on the suffix tree for \(S = \mathrm{abaababbabbab\$}\).}
\label{fig:tree}
\end{center}
\end{figure}

We start by labelling with 0 the root of $\ST_1$ and marking it.  For \(1 \leq i \leq n\), we find the lowest marked ancestor $u$ in $\ST_1$ of the leaf $w$ labelled $i$.  (This is the difference between building a position heap and Bannai et al.'s algorithm for building the LZ78 trie: they consider only values of $i$ that are the starting positions of phrases in the LZ78 parse.)  If $u$ is $w$ itself, then we simply mark it; otherwise, we find the child $v$ of $u$ that is also an ancestor of $w$.  If $u$ has a constant number of children then finding $v$ takes $\Oh{1}$ time even in $\ST_1$.  If $u$ has more than a constant number of children then we use $\ST_2$ to find $v$, as we explain next.  If $v$'s stringdepth (i.e., the length of its path label) is 1 more than $u$'s, then we move the label $i$ to $v$ and mark it in $\ST_1$.  Otherwise, we insert a new node $v'$ between $u$ and $v$ in $\ST_1$; assign the first character of the edge label of the old edge \((u, v)\) to the new edge \((u, v')\) and assign the rest to the new edge \((v', v)\); move the label $i$ to $v'$; and mark $v'$.  This all takes $\Oh{1}$ amortized time.  Finally, for \(1 \leq i \leq n\), we add a maximal-reach pointer from the node labelled $i$ in $\ST_1$ to the deepest marked ancestor of the leaf originally labelled $i$.  Figure~\ref{fig:tree} shows $\heap$ overlaid on the suffix tree for \(S = \mathrm{abaababbabbab\$}\).  In this case, building $\heap$ requires us to insert into $\ST_1$ the nodes of the heap labelled 3, 7, 9 and 10.

Notice that, if $u$ has more than one child then, first, $u$ exists in both $\ST_1$ and $\ST_2$ and, second, we have not inserted any nodes in $u$'s subtree in $\ST_1$.  Therefore, $v$ also exists and is $u$'s child in both $\ST_1$ and $\ST_2$.  We can find $v$ in $\Oh{1}$ time in $\ST_2$ by finding the ancestor of $w$ whose depth is 1 more than $u$'s.  Summing up, we have the following theorem.

\begin{theorem} \label{thm:tree2heap}
Given the suffix tree for a string, we can build the position heap for that string in linear time independent of the size of the alphabet.
\end{theorem}

Since Farach's construction of suffix trees takes linear time independent of the alphabet size, we have answered affirmatively Kucherov's question of whether there is an algorithm for building position heaps that takes linear time independent of the alphabet size.

\begin{corollary} \label{cor:tree2heap}
We can build the position heap for a given string in linear time independent of the size of the alphabet.
\end{corollary}

\section{Using a Position Heap as a Suffix Array} \label{sec:heap2array}

The order in which we see positions in a traversal of $\heap$ may not be the order in which they appear from left to right on the leaves of the suffix tree for $S$, which is the same as their order in the suffix array \(\SA [1..n]\) for $S$.  For example, if \(S = \mathrm{abaababbabbab\$}\) then \(\SA = [14, 3, 12, 1, 4, 9, 6, 13, 2, 11, 8, 5, 10, 7]\); since the node labelled 4 is the child of the node labelled 1 and the parent of the node labelled 12 in $\heap$, no traversal of $\heap$ can produce the order \(12, 1, 4\).  Nevertheless, by the definition of a position heap, if positions are labels of nodes at the same depth in $\heap$, then their left-to-right order is the same as the lexicographic order of the suffixes starting at those positions and, so, the same as their left-to-right order in the suffix tree or suffix array for $S$.

Let \(D [1..n]\) be the array in which \(D [i]\) is the depth in $\heap$ of the value \(\SA [i]\).  In other words, if \(D [i]\) is the $r$th copy of $d$ in $D$, then the label of the $r$th node from the left at depth $d$ in $\heap$ is \(\SA [i]\).  For example, if \(S = \mathrm{abaababbabbab\$}\) then \(D = [1, 2, 3, 1, 2, 4, 3, 2, 1, 4, 3, 2, 3, 2]\).  Since \(D [8]\) is the third copy of 2 in $D$, the label on the third node from the left at depth 2 in $\heap$ is \(\SA [8] = 13\), as shown in Figure~\ref{fig:heap}.  It follows that, if we can answer access and partial rank queries on $D$ and access nodes in $\heap$ given their depths and their ranks from the left at those depths, then we can support access to $\SA$.

We can store $D$ in \(n H_0 (D) + o (n (H_0 (D) + 1))\) bits, where \(H (D) \leq \log h\) is the 0th-order empirical entropy of $D$ and $h$ is the height of $\heap$, such that access and partial rank queries take $\Oh{1}$ time~\cite{BGN12}.  Ehrenfeucht et al. showed that, although $h$ can be as large as $n$ in the worst case, it is typically $\Oh{\log n}$.  There are \((2 n + o (n))\)-bit data structures that support access in $\Oh{1}$ time to any node in $\heap$ given its rank in pre-order, in-order or post-order traversals; given a pointer to a node, they also return its rank in the appropriate traversal.  Notice that any of these traversals visits the nodes at any particular depth in $\heap$ in their left-to-right order.  For the sake of simplicity, we now consider only pre-order traversal.

Let \(E [1..n]\) be the array in which \(E [i]\) is the depth of the \((i + 1)\)st node (or $i$th if we ignore the root) visited in a pre-order traversal of $\heap$.  In other words, if \(E [i]\) is the $r$th copy of $d$ in $E$, then the $r$th node from the left at depth $d$ is the \((i + 1)\)st visited in a pre-order traversal.  For example, if \(S = \mathrm{abaababbabbab\$}\) then \(E = [1, 1, 2, 2, 3, 3, 4, 1, 2, 2, 3, 4, 2, 3]\).  Since \(E [9]\) is the third copy of 2 in $E$, the third node from the left at depth 2 is the 9th node visited in a pre-order traversal of $\heap$.  It follows that, if we can answer select queries on $E$, then we can access nodes in $\heap$ given their depths and their ranks from the left at those depths.

We can store $E$ in \((1 + \epsilon) n H_0 (E) + o (n)\) bits such that select queries take $\Oh{1}$ time~\cite{BCGNN12}, where $\epsilon$ is any positive constant.  Notice that $E$ is a permutation of $D$ so \(H_0 (E) = H_0 (D) \leq \log h\).  Therefore, we can add \((2 + \epsilon) n H_0 (D) + o (n (H_0 (D) + 1)) = \Oh{n \log h}\) bits to $\heap$ and support access to $\SA$ in $\Oh{1}$ time.

The inverse suffix array \(\SA^{- 1} [1..n]\) stores the lexicographic ranks of the suffixes in left-to-right order.  For example, if \(S = \mathrm{abaababbabbab\$}\) then \(\SA^{- 1} = [4, 9, 2, 5, 12, 7, 14, 11, 6, 13, 10, 3, 8, 1]\).  Suppose we store data structures supporting access and partial rank queries on $E$ and select queries on $D$, which take another \((2 + \epsilon) n H_0 (D) + o (n (H_0 (D) + 1)) = \Oh{n \log h}\) bits.  If we want to access \(\SA^{- 1} [i]\), then we follow the pointer to the node $v$ labelled $i$ in $\heap$; find $v$'s rank $t$ in the pre-order traversal of $\heap$; find the partial rank $r$ of \(E [t - 1] = d\) in $E$; and use select to find the position of the $r$th copy of $d$ in $D$.  This takes a total of $\Oh{1}$ time.  For example, to access \(\SA^{- 1} [13]\), we find the node labelled 13 in $\heap$, which is the 10th visited in a pre-order traversal; find the partial rank 3 of \(E [9] = 2\) in $E$; and return the position 8 of the third 2 in $D$.

\begin{theorem} \label{thm:heap2array}
We can add $\Oh{n \log h}$ bits to a position heap, where $h$ is the height of the heap, such that it supports access to the corresponding suffix array and inverse suffix array in $\Oh{1}$ time.
\end{theorem}

\section{Using a Compressed Suffix Array as a Position Heap} \label{sec:array2heap}

Many compressed suffix arrays (see, e.g.,~\cite{NM07} for a survey) support efficient access to both $\SA$ and $\SA^{- 1}$.  Suppose we have access to $\SA$ and $\SA^{- 1}$ and want to represent a position heap, including
\begin{itemize}
\item its structure as a tree;
\item the nodes' labels;
\item the edges' labels;
\item the maximal-reach pointers;
\item an array of pointers such that, given $i$, in $\Oh{1}$ time we can find the node labelled $i$;
\item a data structure such that, given $i$ and $j$, in $\Oh{1}$ time we can determine whether the node labelled $i$ is an ancestor of the node labelled $j$.
\end{itemize}
We can represent the heap's structure as a tree using any of the \((2 n + o (n))\)-bit data structures mentioned in Section~\ref{sec:heap2array}; assume we use the one based on pre-order traversal.  Without increasing the size of the data structure by more than \(o (n)\) bits, we can support queries to determine whether one node is the ancestor of another, given pointers to them.  We now show that, with the data structures for access, partial rank and selection on $D$ and $E$, we can represent the nodes' labels and the array of pointers.

To find the label of a given node $v$, we find $v$'s rank $t$ in the pre-order traversal of $\heap$; find the partial rank $r$ of \(E [t - 1] = d\) in $E$; use select to find the position $p$ of the $r$th copy of $d$ in $D$; and return \(\SA [p]\).  For example, if \(S = \mathrm{abaababbabbab\$}\) and we are asked for the label of the 10th node visited in a pre-order traversal of $\heap$, then we find the partial rank 3 of \(E [9] = 2\); find the position 8 of the third 2 in $D$; and return \(\SA [8] = 13\).

To find a node given its label $i$, we find the position \(\SA^{- 1} [i] = p\) in $\SA$ of $i$; find the partial rank $r$ of \(D [i] = d\) in $D$; find the partial rank \(t - 1\) of the $r$th copy of $d$ in $E$; and return a pointer to the $t$ node visited in a pre-order traversal of $\heap$.  For example, if \(S = \mathrm{abaababbabbab\$}\) and we are asked to find the node in $\heap$ with label 13, then we find the position \(\SA^{- 1} [13] = 8\) in $\SA$ of 13; find the partial rank 3 of \(D [8] = 2\); find the partial rank 9 of the 3rd copy of 2 in $E$; and return a pointer to the 10th node visited in a pre-order traversal of $\heap$.

To be able to return edges' labels, we store a bitvector indicating, for each distinct character $a$, the interval of $\SA$ containing the positions of copies of $a$ in $S$.  Assuming the size $\sigma$ of the alphabet is at most $n$, this bitvector takes \(\sigma \log (n / \sigma) + o (n) = \Oh{n}\) bits, and lets us determine in $\Oh{1}$ time the first character \(S [i]\) in suffix \(S [i..n]\) given \(S [i..n]\)'s lexicographic rank among the suffixes of $S$.  If we are using a compressed suffix array that already supports this functionality, then we do not need the bitvector.

To find an edge's label, we find the label $i$ and depth $d$ of the node at the bottom of that edge, find the position \(\SA^{- 1} [i + d - 1]\) in $\SA$ of \(i + d - 1\), then use the bitvector to determine the character \(S [i + d - 1]\).  For example, if \(S = \mathrm{abaababbabbab\$}\) and we are asked to find the label of the edge above the node labelled 13, which is at depth 2, then we find the position \(\SA^{- 1} [14] = 1\) of 14 in $\SA$ and use the bitvector to determine \(S [14] = \$\).

To be able to return nodes' maximal-reach pointers, we store the balanced-parentheses representation of the tree structure of $\heap$, with copies of a special symbol $*$ interleaved so that the $i$th copy of $*$ occurs after the $j$th copy of `$($' if, in a pre-order traversal of the position heap overlaid on the suffix trie (see Figure~\ref{fig:trie}) we visit the $i$th leaf of the trie after we visit the $j$th node of the heap; the $i$th copy of $*$ occurs before the $j$th copy of `$)$' if, in a post-order traversal of the position heap overlaid on the suffix trie, we visit the $i$th leaf of the trie before visiting the $j$th node of the heap.  For example, if \(S = \mathrm{abaababbabbab\$}\) then we store
\[(\ (\ *\ )\ (\ (\ *\ )\ (\ (\ *\ )\ *\ *\ (\ (\ *\ )\ *\ )\ )\ )\ (\ (\ *\ )\ (\ *\ (\ (\ *\ )\ *\ *\ )\ )\ (\ (\ *\ *\ )\ )\ )\ )\ .\]
To clarify this example, we now attach subscripts and superscripts showing the labels of the nodes of the heap to which parentheses correspond, and superscripts showing the labels of the leaves of the trie to which copies of $*$ correspond:
\begin{eqnarray*}
&& (_0\ (_{14}\ *^{14}\ _{14})\ (_1\ (_3\ *^3\ _3)\ (_4\ (_{12}\ *^{12}\ _{12})\ *^1\ *^4\ (_6\ (_9\ *^{9}\ _9)\ *^6\ _6)\ _4)\ _1) \ldots\\[.5ex]
&& \mbox{} \hspace{3ex} \ldots (_2\ (_{13}\ *^{13}\ _{13})\ (_5\ *^2\ (_8\ (_{11}\ *^{11}\ _{11})\ *^8\ *^5\ _8)\ _5)\ (_7\ (_{10}\ *^{10}\ *^7\ _{10})\ _7)\ _2)\ _0)\ .
\end{eqnarray*}

Recall from Section~\ref{sec:tree2heap} that the maximal-reach pointer of the node labelled $i$ in $\heap$ points to the deepest node of $\heap$ that, when $\heap$ is overlaid on the suffix trie, is an ancestor of the leaf labelled $i$ in the suffix trie.  For example, if \(S = \mathrm{abaababbabbab\$}\), then the node labelled 5 in $\heap$ points to the node labelled 8 (see Figure~\ref{fig:tree}).  It follows that the maximal-reach pointer of the node labelled $i$ is to the node corresponding to the matching pair of parentheses most closely enclosing the \(\SA^{- 1} [i]\)th copy of $*$ in our augmented balanced-parentheses representation of $\heap$.  For example, if \(S = \mathrm{abaababbabbab\$}\), then the \(\SA^{- 1} [5] = 12\)th copy of $*$ is most closely enclosed by the matching pair of parentheses corresponding to the 12th node visited in a pre-order traversal of $\heap$, which is labelled 8.  We can store our augmented representation in $\Oh{n}$ bits such that find this matching pair of parentheses, and the corresponding node, in $\Oh{1}$ time.  Carefully combining this with all the results in this section, we obtain the following theorem.

\begin{theorem} \label{thm:array2heap}
Suppose we have a compressed suffix array that supports access to both the suffix array and inverse suffix array in $\Oh{t}$ time, and the corresponding position heap has height $h$.  Then we can add $\Oh{n \log h}$ bits to the compressed suffix array such that it simulates the position heap with an $\Oh{t}$-factor slowdown.
\end{theorem}

\section{Suffix Heaps} \label{sec:sheap}

\begin{figure}[t]
\begin{center}
\includegraphics[width=30ex]{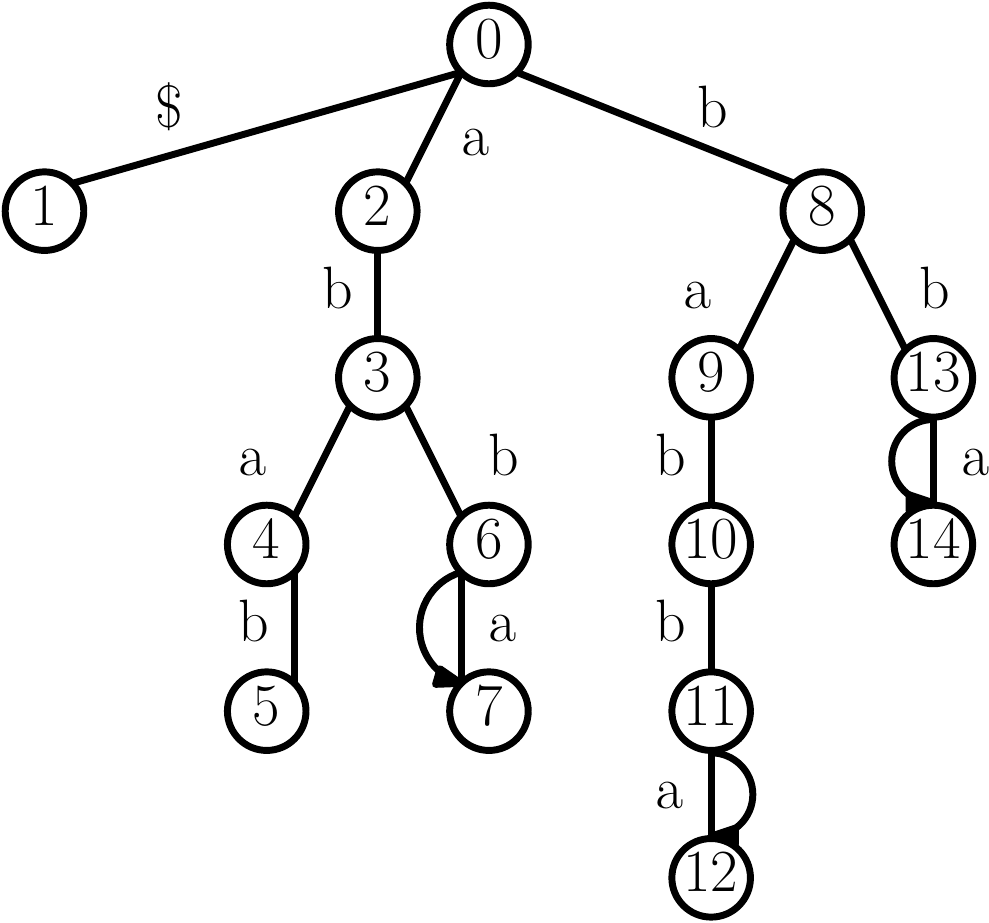}
\caption{The suffix heap $\sheap$ for \(S = \mathrm{abaababbabbab\$}\).}
\label{fig:sheap}
\end{center}
\end{figure}

Suppose we modify the definition of a position heap so that, instead of the path label of the node labelled $i$ being a prefix of \(S [i..n]\), it is a prefix of \(S [\SA [i]..n]\).  We call the resulting data structure the suffix heap $\sheap$ for $S$.  For example, if \(S = \mathrm{abaababbabbab\$}\) then $\sheap$ is as shown in Figure~\ref{fig:sheap} (except that maximal-reach pointers are omitted there when they point back to the nodes themselves).

Searching in a suffix heap is similar to searching in a standard position heap but now, instead of reporting a node's label $i$, we report \(\SA [i]\); instead of computing \(i + d\), we compute \(\SA^{- 1} [\SA [i] + d]\).  Therefore, searching a suffix heap require access to $\SA$ and $\SA^{- 1}$.  For example, to search for \(P = \mathrm{aabab}\) in \(S = \mathrm{abaababbabbab\$}\), we start at the root and descend along the edge labelled \(P [1] = \mathrm{a}\) to the node $v$ labelled 2 at depth 1.  We return to the root and descend to along the edges labelled \(P [2] = \mathrm{a}\), \(P [3] = \mathrm{b}\), \(P [4] = \mathrm{a}\) and \(P [5] = \mathrm{b}\) to the node $v'$ labelled 5.  Since \(\SA^{- 1} [\SA [2] + 1] = 5\) is the label of $v'$, we report position \(\SA [2] = 3\).  We will give more examples in the full version of this paper.

We can build a suffix heap using the linear-time algorithm described in Section~\ref{sec:tree2heap} but first labelling the leaves of the suffix tree by their ranks from left to right.  There is a simpler recursive algorithm, however, to build the suffix heap from the suffix trie; we can make it linear-time by simulating the suffix trie with the suffix tree, as before.  We start by creating the root of $\sheap$; for each child $v$ of the root of the suffix trie, we call \(\build (v, 1)\), where \(\build (v, c)\) is the procedure given in Figure~\ref{alg:build}.  We will prove this algorithm correct and analyze it in the full version of this paper.  If we store trees using their balanced-parentheses representation, then this algorithm takes $\Oh{n}$ bits of work space.

\begin{figure}[t]
\begin{center}
\parbox{55ex}
{create a node $x$\\
let \(\mathit{leaves} (v)\) be the number of leaves in $v$'s subtree\\
if \(c = \mathit{leaves} (v)\) then return $x$\\
let \(v_1, \ldots, v_t\) be $v$'s children\\
for \(i := 1..t\)\\
\mbox{} \hspace{3ex}    if \(\mathit{leaves} (v_i) > c\)\\
\mbox{} \hspace{6ex}        make \(\build (v_i, c + 1)\) a child of $x$\\
\mbox{} \hspace{6ex}        for \(j := i + 1..t\)\\
\mbox{} \hspace{9ex}            make \(\build (v_j, 1)\) a child of $x$\\
\mbox{} \hspace{3ex}    else\\
\mbox{} \hspace{6ex}            \(c := c - \mathit{leaves} (v_i)\)\\
return $x$}
\caption{Pseudocode for the recursive procedure \(\build (v, c)\).}
\label{alg:build}
\end{center}
\end{figure}

Notice that nodes' labels are simply their ranks in a pre-order traversal of $\sheap$; therefore, in a total of \(2 n + o (n)\) bits we can store
\begin{itemize}
\item $\sheap$'s structure as a tree;
\item the nodes' labels'
\item an array of pointers such that, given $i$, in $\Oh{1}$ time we can find the node labelled $i$;
\item a data structure such that, given $i$ and $j$, in $\Oh{1}$ time we can determine whether the node labelled $i$ is an ancestor of the node labelled $j$.
\end{itemize}

Suppose we have stored the bitvector described in Section~\ref{sec:array2heap} that indicates, for each distinct character $a$, the interval of $\SA$ containing the positions of copies of $a$ in $S$.  Again, assuming the size $\sigma$ of the alphabet is at most $n$, this bitvector takes $\Oh{n}$ bits and lets us determine in $\Oh{1}$ time the first character \(S [i]\) in suffix \(S [i..n]\) given \(S [i..n]\)'s lexicographic rank among the suffixes of $S$.  If we are using a compressed suffix array that already supports this functionality, then we do not need the bitvector.

To find an edge's label, we find the label $j$ and depth $d$ of the node at the bottom of that edge; find the starting position \(i = \SA [j]\) in $S$ of the lexicographically $j$th suffix; find the position \(\SA^{- 1} [i + d - 1]\) in $\SA$ of \(\SA [j] + d - 1\); then use the bitvector to determine the character \(S [i + d - 1]\).  For example, if \(S = \mathrm{abaababbabbab\$}\) and we are asked to find the label of the edge above the node labelled 13, which is at depth 2, then we find the starting position \(\SA [13] = 10\) of the lexicographically 13th suffix; find the position \(\SA^{- 1} [11] = 10\) of 11 in $\SA$; and use the bitvector to determine \(S [11] = \mathrm{b}\).

Suppose the maximal-reach pointer of the node labelled $i$ is to the node labelled $j$ at depth $d$.  Then
\begin{eqnarray*}
\lefteqn{S \left[ \rule{0ex}{2ex} \SA [i]..\SA [i] + d - 1 \right]}\\
& = & S \left[ \rule{0ex}{2ex} \SA [i + 1]..\SA [i + 1] + d - 1 \right]\\
&& \hspace{3ex} \vdots\\
& = & S \left[ \rule{0ex}{2ex} \SA [j]..\SA [j] + d - 1 \right]\,.
\end{eqnarray*}
It follows that, if the maximal reach pointer of the node labelled \(i' > i\) is to the node labelled $j'$, then \(j' > j\).  Therefore, we can store the nodes' maximal-reach pointers in $\sheap$ as a balanced-parentheses representation of the tree structure with copies of a special symbol $*$ interleaved so that the $i$th copy of $*$ occurs after the $j$th copy of `$($' if the maximal-reach pointer of the node labelled $i$ is to the node labelled $j$.  For example, if \(S = \mathrm{abaababbabbab\$}\) then we store
\begin{eqnarray*}
&& (_0\ (_1\ *^1\ _1)\ (_2\ *^2\ (_3\ *^3\ (_4\ *^4\ (_5\ *^5\ _5) (_6\ (_7\ *^6\ *^7\ _7)\ _6)\ (_8\ *^8\ (_9\ *^9\ (_{10}\ *^{10}\ \ldots\\[.5ex]
&& \mbox{} \hspace{3ex} \ldots(_{11}\ (_{12}\ *^{11}\ *^{12}\ _{12})\ _{11})\ _{10})\ _9)\ (_{13}\ (_{14}\ *^{13}\ *^{14}\ _{14})\ _{13})\ _8)\ _0)\ ;
\end{eqnarray*}
again, we have show subscripts and superscripts only to clarify the example.  Given a pointer to the node labelled $i$, we can find where its maximal-reach pointer points by using a select query to find the position of the $i$th copy of $*$, using a rank query to find the number of copies of `$($' preceding it, and subtracting 1 for the root.  For example, if \(S = \mathrm{abaababbabbab\$}\) and we want the maximal-reach pointer of the node labelled 6, then we compute \(\mathrm{rank}_( \mathrm{select}_* (6) - 1 = 7\).  We can store our augmented balanced-parentheses representation in $\Oh{n}$ bits such that rank and select queries take $\Oh{1}$ time.  Carefully combining this with all the results in this section, we obtain the following theorem.

\begin{theorem} \label{thm:array2sheap}
Suppose we have a compressed suffix array that supports access to both the suffix array and inverse suffix array in $\Oh{t}$ time.  Then we can add $\Oh{n}$ bits such that it simulates the corresponding suffix heap with an $\Oh{t}$-factor slowdown.
\end{theorem}

\section*{Acknowledgments}

Many thanks to Hideo Bannai, Gregory Kucherov and Gonzalo Navarro for their insightful comments.  The first author also thanks his colleagues at Aalto and the University of Helsinki and his students for helpful discussions.

\bibliographystyle{plain}
\bibliography{heaps}

\end{document}